\documentclass[aps,prl,noshowpacs,floats,10pt,twocolumn]{revtex4}
\usepackage{graphicx}
\usepackage{amsfonts}
\usepackage{euler}
\begin{document}
\def\E{{\bf E}}

\def\P{{\bf P}}

\def\D{{\bf D}}

\def\r{{\bf r}}

\def\dV{{\rm d^3}{\bf r}}

\def\qq{{\bf q}}

\def \div{ {\rm div}\, } \def \grad{ {\rm grad}\, } \def \curl{ {\rm curl}\, }

\def\p{{\bf p}}

\def\rhat{\hat {\bf r}}

\def \dv{\; d^3\rr}

\title{Multiscale Monte Carlo for simple fluids} \author{A. C. Maggs}
\affiliation{Laboratoire de Physico-Chime Th\'eorique, UMR CNRS-ESPCI 7083, 10
  rue Vauquelin, 75231 Paris Cedex 05, France.}
\begin{abstract}
  We introduce a multiscale Monte Carlo algorithm to simulate dense simple
  fluids.  The probability of an update follows a power law distribution in
  its length scale.  The collective motion of clusters of particles requires
  generalization of the Metropolis update rule to impose detailed balance. We
  apply the method to the simulation of a Lennard-Jones fluid and show
  improvements in efficiency over conventional Monte Carlo and molecular
  dynamics, eliminating hydrodynamic slowing down.
\end{abstract}
 \maketitle

 Both molecular dynamics and Monte Carlo are widely used in the study of
 fluids.  The first, which is often preferred for large scale studies, allows
 one to access both static and dynamic information; however use of a large
 time step leads to systematic errors in the thermodynamics \cite{geometric}.
 Monte Carlo has the advantage of converging to the exact equilibrium
 distribution without any systematic errors;  recently the introduction of
 collective updates has led to substantial improvements in the efficiency of
 Monte Carlo simulation of fluids.  Particularly notable is a cluster
 algorithm for hard spheres \cite{krauth} and its generalization 
\cite{luijten2} to Lennard-Jones fluids. These papers introduced 
 updates based on reflection or rotation of groups of particles with respect
 to a randomly chosen centre.  The algorithms are particularly well adapted to
 the simulation of dilute, heterogeneous systems.  They work less well for
 dense homogeneous fluids. In this Letter we introduce collective updates
 which simulate such fluids more efficiently.

 Our method displaces blocks of particles on a scale, $\ell$, which is
 intermediate between the dimensions of the particles and the simulation cell.
 By moving groups of particles we accelerate the dynamics of long wavelength
 fluctuations.  Three technical problems are to be solved in the
 implementation of the method: Firstly we must move particles without
 substantially increasing the energy of the system; otherwise the Metropolis
 algorithm will refuse the trial.  Secondly we create a set of moves that are
 {\it reversible}, in order to apply detailed balance. Thirdly we calculate
 the Jacobian that is implied by our collective moves, which complicates the
 counting of the number of states.  We treat each of these three points before
 presenting an implementation and studying the relaxation of long wavelength
 density fluctuations.

 {\it Rigidly}\/ displacing a block of particles a distance $\epsilon$
 generates a mismatch in the structure of the fluid at the interface between
 the moved and stationary particles. This mismatch rapidly leads to a high
 energy cost and low efficiency. We mitigate the problem by introducing
 updates in which the mismatch is {\it spread over the whole volume}\/
 $\ell^d$ in $d$ dimensions.  We choose updates along a single coordinate direction $\alpha$
\begin{equation}
  r'_{i,{\alpha}} = r_{i,{\alpha}} + \epsilon \,
  g \left( {\r_i-{\bf c}_0 \over \ell} \right), \quad \quad \forall i \label{func}
\end{equation}
for particles, $i$ within $\ell$ of ${\bf c}_0$.  The continuous function
${g}$ has the properties $g(0)=1$, and $g(\r) = 0$ for $|\r| >1$.  $\epsilon$
is a random amplitude, ${\bf c}_0$ a randomly chosen centre in the simulation
volume. Thus only particles within $\ell$ of the point ${\bf c}_0$ move.

The proposed displacement, $\epsilon g$ generates a map $x \rightarrow G(x) $ on the
configuration space of the system, where $x$ designates the $N\times d$
coordinates of the $N$ particles.  The simplest manner of generating dynamics
that converge to the Boltzmann distribution is the imposition of {\it detailed
  balance}\/ between pairs of states that are linked in both directions by a
transformation.  However as yet $G$ {\it is not paired with an inverse
  transformation}\/, $G^{-1}$.  We construct explicitly the inverse.  We
add to the  moves eq.~(\ref{func}) a second set found by solving
eq.~(\ref{func}) for $\r_i$ as a function of $\r'_i$.  We then choose 
 between direct and inverted moves with probability $0.5$.

We now turn to the imposition of detail balance. Standard derivations consider
a discrete space: In the presence of transitions between two states $i$, $j$
with energies $E_i$ and $E_j$ one requires that
\begin{equation}
  p(i\rightarrow j) \pi_i = p(j \rightarrow i) \pi_j
\end{equation}
where the occupation probabilities are given by the Boltzmann weight
$\pi_i=e^{-\beta E_i}/Z$.  One solution for the transition rates
$p(i\rightarrow j)$ is the Metropolis choice:
\begin{equation}
  p( i\rightarrow j ) = \min\left(1, e^{-\beta(E_j-E_i)}\right)
\end{equation}

In the continuum one should count the number states in some neighborhood $dx$
of $x$ and not just work with the probability density at $x$.  Consider two
states $x$, $x'$ linked by the transformations $x'=G(x)$, $x'=G^{-1}(x)$.  The
transformation distorts the the volumes $dx$, and $dx'$ which are related by
the Jacobian $J_{G}= \left |{\partial (x')\over \partial (x)} \right |$.  The
pair $\{G, G^{-1}\}$ bring two volumes $dx$ and $dx/J_{G}$ into
correspondence.  The number of states near $x$ is $\pi_x dx$; near $x'$ there
are $\pi_{x'} dx/J_G $.  The transition probabilities should then satisfy
\begin{equation}
  p(x\rightarrow x') \pi_x =  p (x'\rightarrow x)  {\pi_{x'} \over J_G} 
\end{equation}
in order to generate the probability density $\pi$.  This equation has a
solution
\begin{equation}
  p(x \rightarrow x') = \min 
  \left (1, e^{-\beta (E_{x'} + T \log {J_G}  -E_x )}\right )
  \label{newmetropolis}
  \label{jacobian}
\end{equation}
with $T=\beta^{-1}$ the temperature.  We note that Jacobian weighting factors
are used in Monte Carlo simulations of polymers \cite{torsional} when working
with torsional degrees of freedom.  Standard continuum Monte Carlo updates of
the form $r'_{i,\alpha} = r_{i,\alpha}+\epsilon$ have $J_G=1$ and fall within
our formalism.

The final choices that need to be made concern $\epsilon$ and $\ell$.  The
energy cost of a trial is estimated from the strain induced in the fluid by
the transformation:
\begin{math}
  S \sim \epsilon\, \nabla g( \r/\ell) \sim \epsilon/\ell
\end{math}.  This energy varies as $\int  S^2  \sim S^2 \ell^d \sim \epsilon^2
\ell^{d-2} $.  We match this to the thermal energy and choose
\begin{equation}
  \langle \epsilon^2 \rangle \sim {T\over \ell^{d-2}} \label{epsilon}
\end{equation}
setting the scale of the random amplitude $\epsilon$ \footnote { On repeating the
  argument with a block of rigidly displaced particles we find the exponent $d-2$ 
  replaced by $d-1$. Large blocks can not move as far.}.
We sample $\ell$ from a distribution: Standard arguments give the density of
states in Fourier space as $N(q) dq \sim q^{d-1} dq$.  When this is expressed
as a function of a length scale $\ell=1/q$ we find $N(\ell) d\ell =
d\ell/\ell^{d+1}$. We cut off the distribution at short distances $\ell < l_c$
and sample $\ell$ with probability
\begin{equation}
  P(\ell) \sim {l_c^d\over   \ell^{d+1}}, \quad \quad l_c<\ell<L/2
  \label{prob}
\end{equation}
Fluctuations at all length scales are then sampled at the same rate.

The asymptotic cost of the algorithm per trial can be estimated by recognizing
that the effort needed to update the scale $\ell$ is $O(\ell^d)$ when
interactions are short ranged. The average work per collective update is then
$ \int \ell^d P(\ell) d \ell \sim l_c^d\log{(L/l_c)}$, where $L$ is the system
size.  The effort needed to perform the collective updates diverges only
weakly with system size.

We implemented the algorithm, simulating Lennard-Jones
particles with the potential
\begin{equation}
  U = 4 \epsilon_{LJ} \left [ \left ( \sigma \over r \right )^{12} - 
    \left ( \sigma \over   r \right )^6 \right ]
\end{equation}
truncating the interactions at a radius of $2.5 \sigma$ and shifting $U$ by a
constant to eliminate the discontinuity \cite{smit}. We use length units of
$\sigma$ and energy units of $\epsilon_{LJ}$.  Temperature is measured in
units of $\epsilon_{LJ}$.  Periodic boundary conditions were imposed.

A convenient choice for $g$ is
\begin{equation}
  g(\r/\ell) = 1-r^2/\ell^2
  \label{gx}
\end{equation}
for $r<\ell$, $g=0$ otherwise. This allows the direct inversion of
eq.~(\ref{func}). However, we also implemented  more elaborate
expressions for which analytic inversion was not possible. For these cases we
performed a numerical inversion of eq.~({\ref{func}}) using Newton iteration.
Even here generating the trial moves was a minor contribution to the total
computational cost of the algorithm.

\begin{figure}[htb]
  \includegraphics[scale=.48] {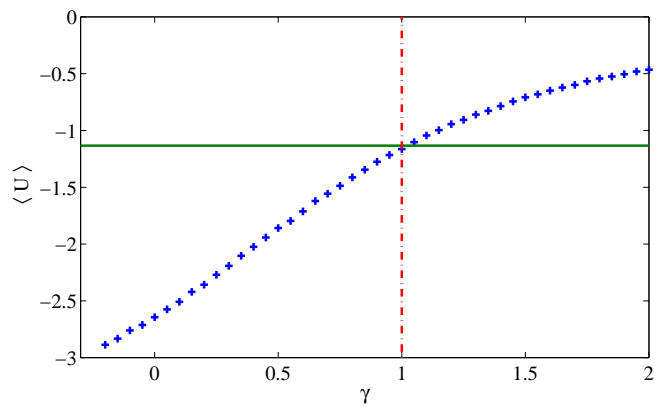}
\caption{Variation of the mean energy, $\langle U \rangle$,
  as a function of the prefactor of $\log{J_G}$. The correct average energy
  is generated with $\gamma=1$.  Errors result if one neglects the Jacobian,
  $\gamma=0$.  Solid line: local Monte Carlo.  Points: multiscale Monte Carlo.
  Statistical errors small than symbol size.  Vertical line guide to eye for
  $\gamma=1$.
\label{fig1}
}\end{figure}

The Jacobian is calculated as a product of individual particle contributions.
For a displacement in the direction $\alpha$
\begin{equation}
  J_G = \prod_i (1 + \epsilon \partial g(\r_i)/\partial r_{i,\alpha})
\end{equation}
The first simulations were performed on a very small system in order to check
the correctness of our arguments as to the need for a Jacobian in the
Metropolis criterion.  To do this we multiplied the term in $\log {J_G}$ by a
numerical prefactor $\gamma$.  $\gamma=1$ corresponds to the algorithm as
described above, $\gamma=0$ corresponds to neglecting the difference between
the volumes $dx$ and $dx'$.  We took five particles in a two dimensional box
of side $L=10$ and simulated using a conventional Monte Carlo algorithm as
well as the multiscale algorithm. As can be seen from Figure~\ref{fig1} use of
the correct Jacobian is essential. These trials on small systems allowed the
generation of high statistics runs, from which we checked the 
correctness of the code using both the energy and partial structure factors.

In order to study the efficiency of the algorithm 
we simulated a two-dimensional fluid with cell size $L=60$, $80$, $120$ and
$180$ at a density $\rho=0.755$, $T=0.5$. 
We performed simulations with conventional,
single particle Monte Carlo, molecular dynamics and multiscale Monte Carlo.
In the simulations we measured the static structure factor
\begin{math}
  S_q= \left|{1\over N}\sum_i e^{i {\bf q.r_i}}\right|^2
\end{math}. This correlation function is used to calculate the compressibility
for $q\rightarrow 0$.  In charged systems similar, long wavelength,
polarization correlation functions must be measured with high accuracy in
order to deduce the dielectric constant \cite{kirkwood}.  $S_q$ is slow to
equilibrate at small $q$ due to the locality of particle motion in
conventional Monte Carlo: A hydrodynamic description of the dynamics involves
a local conserved quantity, the density, which diffuses: {\it i.e.}\/ Model B
dynamics \cite{halperin}. Relaxation rates vary as $\Gamma_q = D q^2$, with
$D$ the collective diffusion coefficient. Similarly in molecular dynamics
propagating sound waves imply that $\Gamma^2_q = c^2 q^2$.

Slow long wavelength fluctuations have consequences for the equilibration of
the energy.  For Monte Carlo the autocorrelation of the energy for the mode
$q$ can be expressed in the form $C_q(t) = \tilde C(Dq^2t)$ with $\tilde C(0)=O(T^2)$, so that in $d$ dimensions
the two time energy-energy correlation function is given by $C_u = \int \tilde
C (D q^2
t)\, {\rm d}^dq$.  One finds
\begin{math}
  C_u(t)\sim 
  {1/ t^{d/2}}
\end{math} \footnote{We also checked this decay in two dimensions using
  Kawasaki dynamics for a lattice gas.}.
It is the integral of $C_u$ that determines the convergence rate of the
average energy \cite{blocking}.  In two dimensions, where $\int C_u(t)\, dt$
diverges for large $t$, short ranged functions converge slowly due to
hydrodynamics. Good thermodynamic statistics require that all modes in the
sample are equilibrated. The situation is better in three dimensions, the
integral is dominated by short wavelength fluctuations.

\begin{figure}[htb]
  \includegraphics[scale=.58] {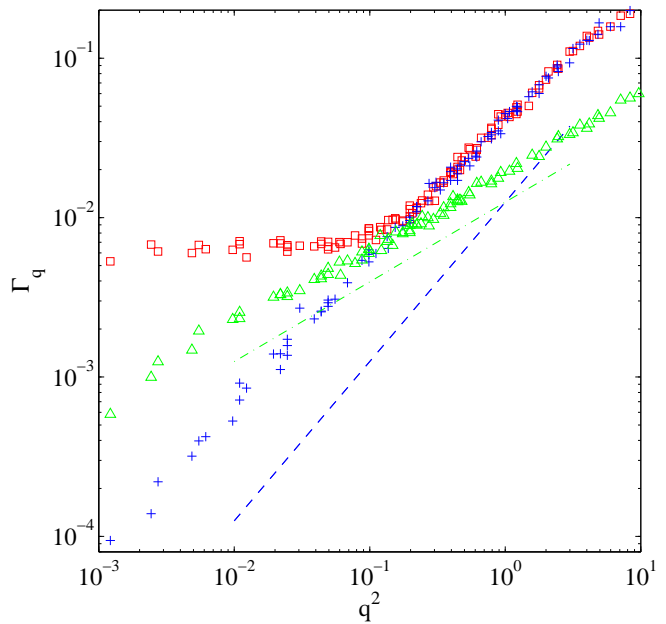}
\caption{Inverse integrated autocorrelation time in sweeps, $\Gamma_q$, 
  for $S_q$.  Conventional  Monte Carlo: $+$. Multiscale: $\square$.
   Molecular dynamics: $\triangle$.  Conventional algorithm has slow, diffusive
  modes.  Multiscale has decay rate independent of $q$ at long wavelengths.
  Dashed line: $\Gamma\sim q^2$; dot-dashed line $\Gamma\sim q$.  Time scale
  for Monte Carlo inverse sweeps. System sizes from $L=40$ to $L=180$.
  Simulations of $2^{20} \approx 10^6$ sweeps, taking up to two weeks of simulation time
  for the largest systems.
\label{fig2}
}\end{figure}

We verified that the algorithm worked correctly when only using updates from
the multiscale distribution. However this turns out to be sub-optimal for the
performance of the simulation. Standard Monte Carlo methods work very well at
the scale of single particles; it is only at larger distances that they become
slow. We  stochastically mixed a standard Monte Carlo algorithm with the
multiscale method using a lower cut off $l_c=7.5$. We used a proportion of
$1000$ local updates for each multiscale update. With these choices of
parameters the execution time of the program increases approximately $30\%$
for $L=40$ and $60\%$ for $L=180$ compared with the purely local code.
Acceptance rates of all moves were adjusted to be approximately $40\%$.  We
verified that the acceptance rate of multiscale moves was almost independent
of $\ell$ by binning acceptance.  We determined the integrated autocorrelation
times of $S_q$ with a blocking algorithm \cite{blocking} and plot the inverse
time, $\Gamma_q$ in Figure~(\ref{fig2}).  As expected local Monte Carlo leads
to slow relaxation of long wavelength modes, with $\Gamma_q = D q^2$.

For the molecular dynamics simulations we used a time step $\tau=0.006$ and a
damping coefficient for the Langevin thermostat of $0.2$. The time step is
typical for low accuracy molecular dynamics studies, but larger than that
which must be used for accurate studies of thermodynamic properties, for
example $\tau=.0014$ in \cite{fisher}.  The dynamics (in Fourier space) of a
system of particles coupled to an external thermostat lead to a dispersion law
familiar from the damped harmonic oscillator \footnote{We neglect internal
  friction of the fluid which leads to weaker dampening.}:
\begin{math}
K q^2 - i \eta \omega  - \rho \omega^2=0
\end{math}
with $\eta$ a friction coefficient, $K$ a compressibility and $\rho$ the
mass density.  For large $q$ modes are under-damped so that $\omega^2
=K q^2/\rho$, corresponding to propagating sound waves. At small $q$
we crossover to an over-damped  "Brownian" regime where $\omega=i K
q^2/\eta$.  
In our simulations we placed the crossover near $q^2=10^{-2}$ by performing
several trial runs with different $\eta$. In this way we ensure propagative
behavior over most of the range of the curve.  At short length scales
molecular dynamics is slower than local Monte Carlo,
however it does a better job of relaxing long wavelength correlations.  We
adjusted the number of time steps in the molecular dynamics simulation so that
the total simulation time was identical to the conventional Monte Carlo code.

The multiscale results are again displayed with a time scale of inverse
sweeps, for a ``clock time'' comparison one must lower the values of $\Gamma$ by
a factor between $1.3$ and $1.6$ depending on $L$.  The multiscale algorithm
displays two regimes. At short wavelengths local Monte Carlo is active leading
to standard $q^2$ behavior. At long wavelengths the algorithm eliminates
hydrodynamic slowing down- all long wavelength modes relax at very similar
rates. For $L=180$ the multiscale algorithm relaxes long wavelength modes $40$
times faster (clock time) than conventional local Monte Carlo. Multiscale is
also faster than molecular dynamics at this scale.

Given the success of the algorithm we tried to find other updates which couple
even more strongly to density fluctuations. We implemented a trial update in
which motion is purely radial about a random centre ${\bf c}_0$. If ${\bf r}$ denotes the
position of a particle with respect to ${\bf c}_0$ we tried
\begin{equation}
  \r' = \r + \epsilon\, \r ( 1 -r/\ell) \label{radial}
\end{equation}
for $r/\ell <1$, $\r'=\r$ otherwise. The results were disappointing. Similar
mediocre results were found for purely rotational updates of the form
\begin{equation}
  \theta' = \theta  + \epsilon \, (1-r^2/\ell^2) \label{rotational}
\end{equation}
with $\theta$ the angular position of a particle seen from ${\bf c}_0$.
Mixtures of eq.~(\ref{radial}) and eq.~(\ref{rotational}) were just as
poor.  Clearly, not all multiscale updates are created equal.

\begin{figure}[htb]
  \includegraphics[scale=.58] {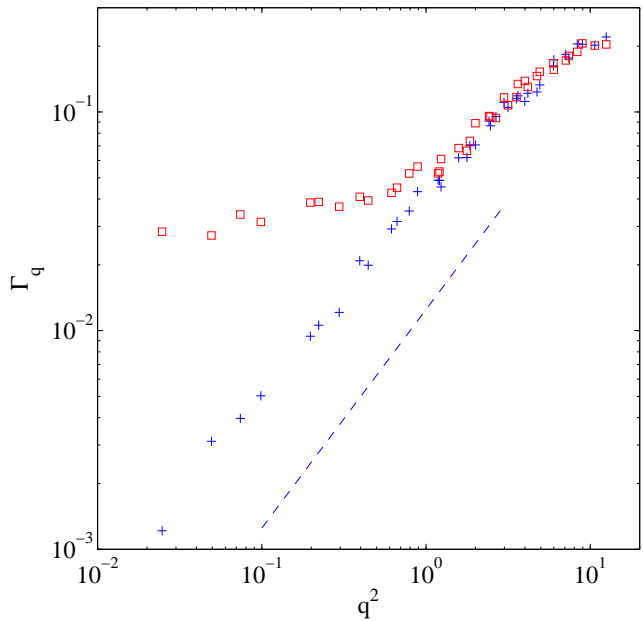}
\caption{Inverse integrated autocorrelation time in sweeps, $\Gamma_q$, 
  for $S_q$ in three dimensions.   Conventional  Monte Carlo: $+$. Multiscale: $\square$.
	$T=1$, $L=40$, $l_c=2.5$. Line $ \Gamma \sim q^2$.
$\Gamma_q$ tends to a constant at long wavelengths for the multiscale algorithm.
\label{fig3}
}\end{figure}

Whilst we performed the most detailed simulations in two dimensions we did
implement the three dimensional version of the code and used it to simulate a
single system with $L=40$. We performed both conventional and multiscale Monte Carlo
and found, Figure \ref{fig3}, qualitatively similar results to  Figure \ref{fig2}.
It is to be noted that three dimensional simulations require considerably
more computing resources for a given system size. The number of particles
is larger, as well as the number of neighboring particles within $2.5 \sigma$. 

In this Letter we have introduced a wider choice of Monte Carlo moves for the
simulation of simple fluids.  A good choice of updates leads to increases in
the efficiency of simulations in a manner reminiscent of multigrid Monte Carlo
simulation of lattice models \cite{sokal}.  Using conventional Monte Carlo we
demonstrated that the relaxation rate varies as $\Gamma=D q^2$; density
fluctuations in large systems take $O(L^2)$ sweeps to equilibrate.  Multiscale
updates, with an average cost of $l_c^d\log{L/l_c}$ per update, eliminate long
wavelength slowing down, leading to an algorithm which is asymptotically
faster.

Our method  permits more general implementations of preferential sampling
\cite{scheraga} in which updates are performed more often in the neighborhood
of an interesting site or surface. It may also have application in
heterogeneous colloidal systems where co-motion of small and large particles
or molecules may be an alternative to cluster methods at high densities.

\bibliography{polar}
\end{document}